\documentclass[11pt]{article}
\usepackage[T1]{fontenc}
\usepackage[utf8]{inputenc}
\usepackage{lmodern}
\usepackage[a4paper,margin=2.6cm]{geometry}
\usepackage{microtype}
\usepackage{amsmath,amssymb}
\usepackage{graphicx}
\usepackage{booktabs}
\usepackage{seqsplit}
\usepackage{tabularx}
\usepackage[round]{natbib}
\usepackage{url}
\usepackage{xcolor}
\usepackage[colorlinks=true,linkcolor=blue!50!black,citecolor=blue!50!black,urlcolor=blue!50!black]{hyperref}
\usepackage{setspace}
\title{Deceleration without identification: testing a power-law trend in Bitcoin prices under persistent noise}
\author{Carlos Baquero\\\small DEI, Faculdade de Engenharia, Universidade do Porto
  \and Raquel Menezes\\\small Departamento de Matemática, Universidade do Minho}
\date{}

\begin{document}
\maketitle

\begin{abstract}
\noindent Plotted against time since its first block on logarithmic axes, the price of Bitcoin lies close to a straight line, and a power law in time has been proposed as its long-run trend. A good fit is weak evidence: when deviations from a trend are as persistent as a random walk, regressions on time find trends and high $R^2$ where none exist. We first bound what fifteen years of weekly prices can reveal, then test only what is answerable. Using a few low-frequency averages, invariant to the noise scale and calibrated over a declared family of persistent deviations with a four-year cycle and declining volatility, we test constant drift in expected log price against an added log-age term. The procedure was fixed, validated and publicly hashed before it was applied once. Constant drift is rejected under this family at nominal 5\%, and a log-age coefficient of 5.7, close to the published exponent, is not rejected. The rejection does not survive allowing persistence beyond a random walk, and a Gompertz curve with a ceiling fits about as well as the power law. Whether deviations revert to the trend cannot be decided. Registered forecasts for 2027--2030 will be scored by future prices.
\end{abstract}

\noindent\textbf{Keywords:} Bitcoin; low-frequency inference; persistent noise; power law; pre-registration; trend testing

\section{Introduction}\label{sec:intro}

\begin{figure}[t]
  \centering
  \includegraphics[width=\textwidth]{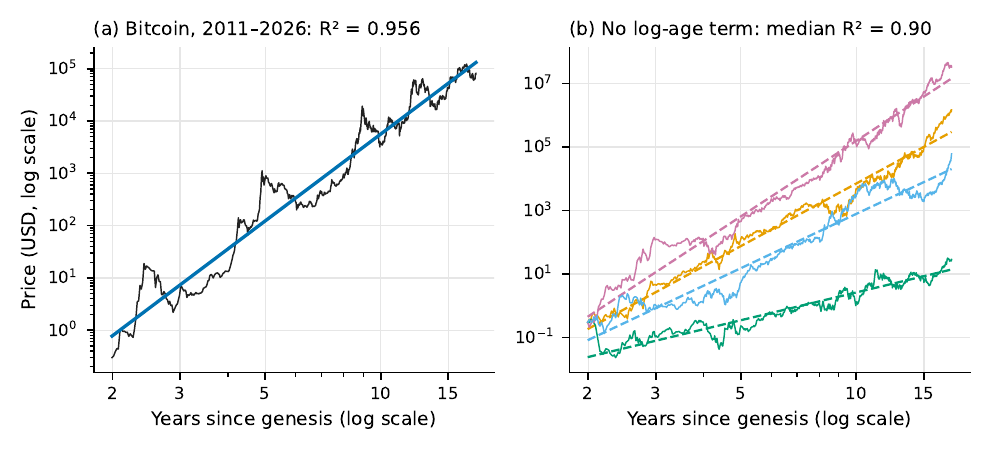}
  \caption{The apparent power law and the same picture from histories without one. (a) Weekly
  closing price of Bitcoin, 2011--2026 (820 Saturdays), against years since the genesis block,
  both on logarithmic scales, with the least-squares power law ($R^2 = 0.956$, exponent 5.51).
  (b) Four simulated histories with constant drift equal to Bitcoin's average weekly log growth,
  random-walk deviations with Bitcoin's volatility profile, and no log-age component. Dashed lines
  are the same least-squares fit. Over 1{,}000 such histories the median $R^2$ is 0.897 (90\%
  range 0.70--0.97), 14\% exceed Bitcoin's $R^2$, and the fitted exponents have median 6.1 (90\%
  range 3.4--9.2). The plotted histories are the first four of the 1{,}000.}
  \label{fig:apparent}
\end{figure}

Plotted against time since its first block, on logarithmic axes, the price of Bitcoin lies close
to a straight line (Figure~\ref{fig:apparent}a). A straight line on these axes is a power law in
time: price grows in proportion to age raised to a fixed exponent. \citet{santostasi2026} fit this
law to daily prices from 2010 to 2026. They report an exponent of $5.69 \pm 0.05$, with
$R^2 = 0.961$. They derive the exponent from two mechanisms, the growth of addresses on a
scale-free network and a Metcalfe-type law relating network size to value. From unit-root and
stationarity tests on the residuals, they conclude that the relation is trend-stationary:
deviations from the line are cyclical and return to it. The model is also discussed in public
commentary as a guide to Bitcoin's long-run value \citep{perrenod2026}.

Which trend describes the record matters to anyone who extrapolates it. A power law implies that
the growth rate falls steadily with age. Constant drift in log price implies that it does not. A
logistic or Gompertz curve implies that growth stops at a ceiling. These curves can describe the
past almost equally well. In our descriptive fits, four of five trend models lie within two
log-likelihood units of one another (Section~\ref{sec:results}). Yet their central forecasts for the end of 2030
range from about 85{,}000 to about 1{,}000{,}000 US dollars (Section~\ref{sec:settle}). Valuations, holdings and
risk limits that rely on trend extrapolation inherit this uncertainty. The statistical question is
therefore not whether a power law fits, but how much the record can tell us about the shape of its
trend.

A good fit is weak evidence on this point, for two reasons. The first is familiar from the study
of power-law distributions. \citet{clauset2009} show that a straight line on a log--log plot is a
necessary but not a sufficient condition. Least-squares fits can give inaccurate exponents, with
no indication of whether the law holds. They propose maximum-likelihood estimation, a
goodness-of-fit test and a comparison with alternative laws. On well-known data sets, their recipe
ruled the power law out for some. For many others it could not separate the power law from
alternatives such as the lognormal. \citet{stumpf2012} add that a statistically sound power law
still needs a mechanism, and that without one results should be reported neutrally.
\citet{gerlach2019} show that dependence between observations reduces their effective number,
which makes standard testing recipes over-confident.

The second reason is specific to time series, and here it is decisive. Deviations of Bitcoin's
price from the fitted power law are highly persistent. Their Durbin--Watson statistic is about
0.005 \citep{santostasi2026}. When deviations wander like a random walk, a regression on time finds
a trend whether or not one exists. Its $R^2$ does not vanish as the sample grows, and its $F$
statistic diverges \citep{durlauf1988}. Its residuals show cycles that the detrending itself
created \citep{nelson1981}. Those results concern linear trends; Figure~\ref{fig:apparent}b shows
the same effect for a power law. Each history there has constant drift and random-walk deviations,
at Bitcoin's average growth and volatility, and no power-law component at all. Fitted exactly as in
Figure~\ref{fig:apparent}a, their median $R^2$ is 0.90, 14\% of them fit better than Bitcoin, and
their fitted exponents centre on 6.1. Regression on log age is non-standard in any case, because
the intercept and log age are asymptotically collinear \citep{phillips2007}. Our record spans only
2.2 units of log age.

We therefore ask first what this record can say, and then test only what it can. We declare a
broad family of models for the deviations. It allows persistence up to a random walk, a stochastic
cycle of about four years, and the decline in volatility seen since 2011. Within it we bound the
power of any test that uses the same low-frequency information (Section~\ref{sec:identifiability}). The trend question is
answerable, but only just. Whether deviations revert to the trend can barely be addressed: no such
test exceeds a power of about 0.2. We then build a test for the answerable part. Does expected log
price grow by more than a constant drift, in the direction of a log-age term? The test uses a few
low-frequency features of the series \citep{muller2008}. It does not depend on the unknown scale of
the noise, and it is calibrated by simulation over the whole declared family (Section~\ref{sec:method}). We fixed
the test, its critical value and its validation gates before applying it, and ran it once
(Section~\ref{sec:freeze}). Because the authors had studied this record before, we present the study as a
pre-registered reanalysis, not as an untouched confirmatory test.

Under the declared family, the test rejects constant drift in expected log price, in the direction
of a positive log-age component. Its validated power at the published exponent was only about
26\%. The rejection is not robust to allowing deviations more persistent than a random walk
(fractional persistence up to $d = 1.4$). The data do not single out the power law: a Gompertz
curve and a power law with added drift have close descriptive likelihoods. Mean reversion remains
untested. We registered forecasts from five trend models for 2027 to 2030, which future prices will
score (Section~\ref{sec:settle}). We also estimate how quickly a longer record sharpens each question
(Section~\ref{sec:discussion}).

\section{Data, trend models and hypotheses}\label{sec:question}

We use one price series, built for this study from exchange records. It takes the closing price
of each UTC day: from Mt.\ Gox trades until the end of 2012, and from Bitstamp's official daily
closes from 2013. We sample it every Saturday from 1 January 2011 to 19 September 2026. One
Saturday has no close and is dropped, which leaves 820 weekly observations. Age is measured in
weeks from the genesis block, mined on 3 January 2009, to the end of each sampled day. The
response $y_t$ is the natural logarithm of the closing price in US dollars. Other price sources,
other weekdays, monthly sampling and inflation-adjusted prices enter as sensitivity analyses
(Section~\ref{sec:results}); the construction and its checks are in Appendix~\ref{app:data}.

The object of every claim in this paper is the trend: the expected log price,
$m(s_t) = \mathrm{E}(y_t)$, as a function of age $s_t$. The deviations $u_t = y_t - m(s_t)$ are
everything else, including bubbles, crashes and cycles. Statements about the trend are statements
about $m$, not about any single price. In particular, a trend can decelerate while prices still
swing by large factors around it.

\begin{figure}[t]
  \centering
  \includegraphics[width=\textwidth]{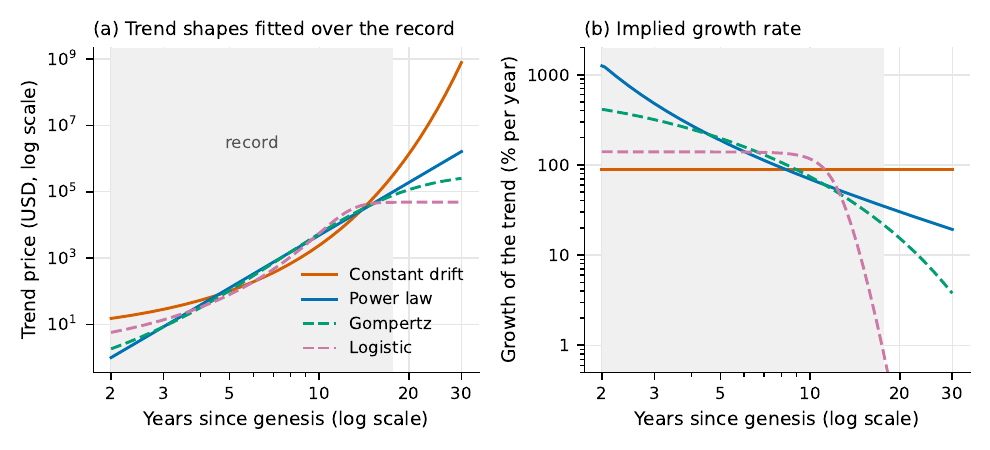}
  \caption{What each trend implies (illustrative curves, no price data). (a) A power law through
  1 US dollar at 2 years and 100{,}000 US dollars at 17.7 years of age (exponent 5.28), with
  constant-drift, Gompertz and logistic curves fitted to it by least squares over the record
  window (shaded). Within the window they stay within 0.6 (Gompertz), 1.7 (logistic) and 2.7
  (constant drift) log units of the power law. At 25 years their trends are about $3\times10^7$,
  $6\times10^5$, $2\times10^5$ and $5\times10^4$ dollars (constant drift, power law, Gompertz,
  logistic). (b) The growth rate each trend implies, in percent per year: constant under constant
  drift; falling like $1/s$ under the power law; falling to zero under the Gompertz and logistic
  curves.}
  \label{fig:shapes}
\end{figure}

Table~\ref{tab:trends} lists the trend models we consider, and Figure~\ref{fig:shapes} shows what
they imply. Under constant drift, log price grows linearly in age, so the trend grows at a fixed
rate. Under a power law, log price is linear in log age, so its growth rate falls in proportion to
$1/s$: the trend decelerates but never stops. Logistic and Gompertz curves approach a ceiling, and
their growth rate falls to zero. Over a record of fixed length these shapes can be hard to tell
apart, yet they imply very different futures (Figure~\ref{fig:shapes}a). We consider no model with
more than three trend parameters, because the low-frequency information that separates trends
supports only a few (Section~\ref{sec:identifiability}).

\begin{table}[t]
  \centering
  \caption{Trend models for expected log price. Age $s$ is measured from the genesis block. Growth
  is the rate of change of the trend, $\mathrm{d}m/\mathrm{d}s$.}
  \label{tab:trends}
  \begin{tabular}{@{}llcl@{}}
    \toprule
    Model & Trend $m(s)$ & Parameters & Growth of the trend \\
    \midrule
    Constant drift & $a + c\,s$ & 2 & constant, $c$ \\
    Power law & $a + b \log s$ & 2 & $b/s$, falls like $1/s$ \\
    Drift and log age & $a + c\,s + b \log s$ & 3 & $c + b/s$, falls towards $c$ \\
    Logistic & $\log K - \log\{1 + e^{-r(s - s_0)}\}$ & 3 & falls to zero (ceiling $K$) \\
    Gompertz & $\log K - \beta e^{-r s}$ & 3 & falls to zero (ceiling $K$) \\
    \bottomrule
  \end{tabular}
\end{table}

The power-law claim has two parts, and we treat them differently. The first concerns the trend:
does expected log price contain a log-age component, so that its growth decelerates? We address it
within the drift-and-log-age model,
\begin{equation}\label{eq:encompassing}
  y_t = a + c\,s_t + b \log s_t + u_t ,
\end{equation}
which contains both constant drift ($b = 0$) and the power law ($c = 0$). Our primary test asks
whether $b > 0$ when $c$ is left free. Leaving the drift free matters. With $c$ fixed at zero, the
test would compare the power law with no growth at all, and any growth would count in its favour. Two secondary checks ask whether the power law needs
an added drift or a curvature in log age; they are not confirmatory. The logistic and Gompertz
curves are compared only descriptively, because the data cannot calibrate a test between them and
the power law (Appendix~\ref{app:fits}). The second part of the claim concerns the deviations: are they
stationary, so that price returns to the trend? Section~\ref{sec:identifiability} shows that this record can barely
address that question, so we report its limits instead of a test.

What makes these questions hard is the behaviour of the deviations. Three features matter. They
are persistent: they can wander for years, which, as Section~\ref{sec:intro} showed, produces trends where there
are none. They contain a cycle of roughly four years, which can mimic curvature over a short
record. And their volatility has fallen since 2011, so a random walk with shrinking steps can look
like a decelerating trend. A test of the trend must hold its error rate under all three. The next
section declares a family of models with these features and asks how much any test could learn
within it.

\section{Identifiability and attainable power}\label{sec:identifiability}

Before testing anything, we ask what the record could reveal under the best possible test. The
answer depends on the deviations, so we first declare which deviation models we allow. The family
has three ingredients, chosen to cover the features described in Section~\ref{sec:question}.

\begin{figure}[t]
  \centering
  \includegraphics[width=\textwidth]{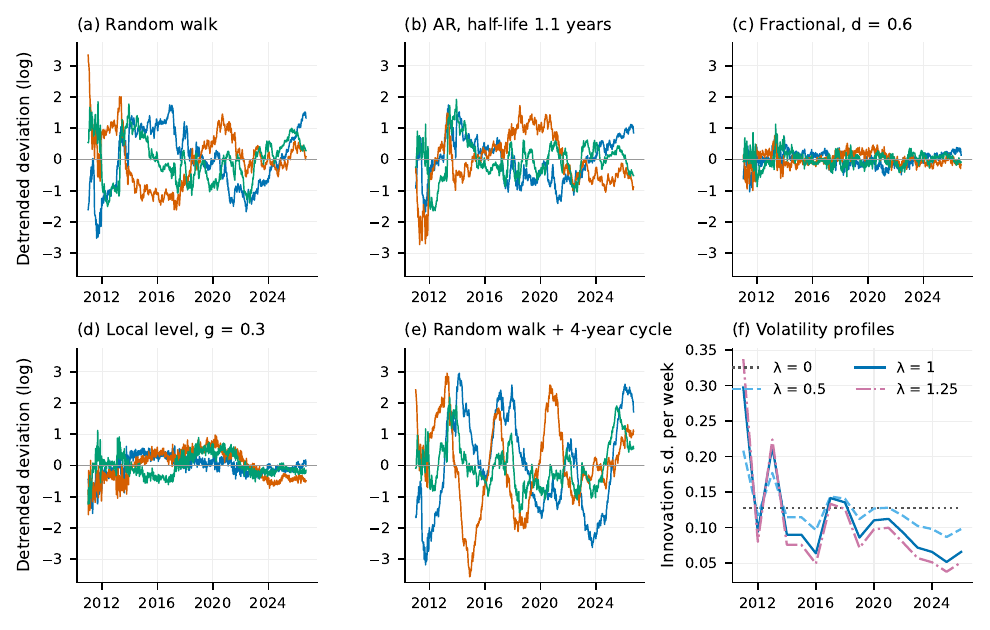}
  \caption{What the declared deviation models look like (simulation, no price data). (a)--(e)
  Three simulated deviation paths for five members of the family, at the 820 weekly dates of the
  record, each shown after removing its own least-squares line in time. All panels use the same
  random numbers, so the differences come from the model alone. (b) Autoregressive with half-life
  1.1 years. (c) Fractional with $d = 0.6$. (d) Random walk plus independent noise. (e) Random walk
  plus a four-year stochastic cycle with standard deviation 0.75 log units. Panels (a)--(e) use
  volatility shape $\lambda = 1$. (f) Weekly innovation standard deviation by year for four
  volatility shapes, each scaled to the same average.}
  \label{fig:noise}
\end{figure}

The first is persistence, in three forms. An autoregressive component reverts to zero with a
half-life from about three months to over twenty years, or not at all: its limiting case is a
random walk. A fractional component decays more slowly than any autoregression, with memory parameter $d$
from 0.25 to 0.9 (Table~\ref{tab:memory} explains the scale of $d$). A local-level component is a random walk observed with independent noise. The
second ingredient is a stochastic cycle with a period of 3.5 to 4.5 years. Its standard deviation
can reach 1.5 log units, and its shape may repeat regularly or drift from one cycle to the next. The
third is the decline in volatility. The innovation scale follows the annual volatility of weekly
returns raised to a power $\lambda$ between 0 (constant) and 1.25. The overall noise scale is left
free. Figure~\ref{fig:noise} shows typical paths. The family is deliberately broad, but it has
limits: persistence stops at a random walk, and the volatility profile has one free parameter.
Section~\ref{sec:results} reports what happens beyond both.

\begin{table}[t]
  \centering
  \caption{What the memory parameter $d$ of a fractional deviation means. The declared family uses
  $d$ from 0.25 to 0.9, plus the random walk; one sensitivity analysis extends it to 1.4. Memory
  concerns how long a shock lasts, not how large it is: the declining size of shocks is described
  separately, by the volatility shape $\lambda$.}
  \label{tab:memory}
  \small
  \begin{tabular}{@{}ll@{}}
    \toprule
    $d$ & What the deviations do \\
    \midrule
    0 & No memory: a shock fades at once \\
    0 to 1 & Long memory: shocks fade, but very slowly \\
    1 & Random walk: a shock never fades, and nothing pulls the path back \\
    above 1 (up to 1.4) & The changes themselves persist, like momentum lasting years, \\
     & so the path makes smooth swings over many years \\
    \bottomrule
  \end{tabular}
\end{table}

Within this family, we summarise each record by a few weighted averages that capture its slowest
movements, following \citet{muller2008}. With 15.7 years of data we use six, whose weights complete
between half a cycle and three cycles over the record. A trend lives in these movements, but so do
persistent deviations; faster fluctuations carry information about volatility, not about the
trend. We also use only the shape of these six numbers, not their size, so that the unknown noise
scale cannot matter. Section~\ref{sec:method} gives the details.

The yardstick is an oracle bound. Take one alternative: a power-law trend with the published
exponent, plus one deviation model. Take one null: constant drift, plus any deviation model in the
family. For this pair, the Neyman--Pearson test is the most powerful test with error rate 5\%.
Any test that holds 5\% across the whole family can do no better against the alternative than
this test does against the hardest null member. The smallest such power, over the null members,
is therefore an upper bound on the power of every valid test that uses the same information. It
is not attainable in general: the oracle knows the alternative exactly.

\begin{figure}[t]
  \centering
  \includegraphics[width=\textwidth]{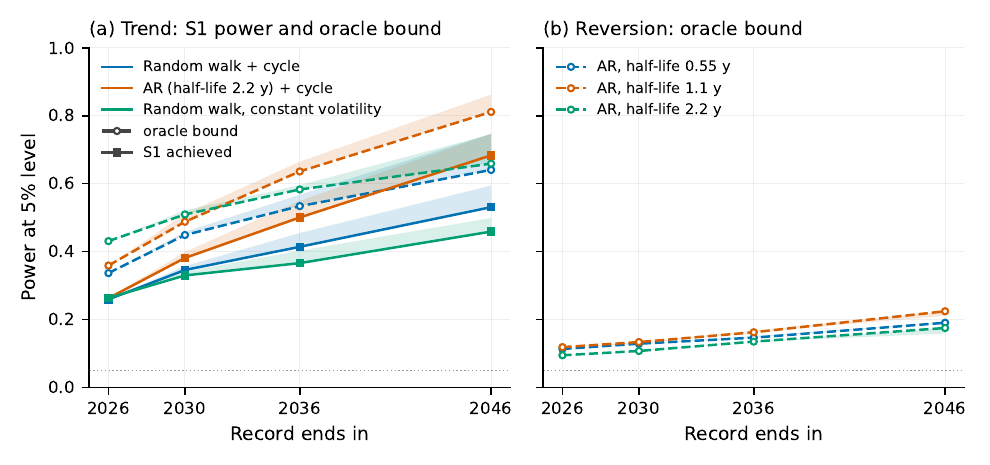}
  \caption{What more years buy (simulation, no price data). (a) Trend question: oracle upper bound
  (dashed, open circles) and the achieved power of our test S1 (solid, squares) at the published
  exponent, for three deviation models, against the year the record ends. Lines assume future
  volatility stays at its 2026 level; shaded bands extend to the case where it keeps falling at its
  past rate. For 2026 the achieved power is from the validation runs; for later years each record
  has its own critical value, set by the same rule. (b) Reversion question: oracle upper bounds for
  autoregressive deviations with half-lives of 0.55, 1.1 and 2.2 years (with a cycle) against
  non-reverting null models. The dotted line marks 5\%, the power of a test that ignores the data.}
  \label{fig:length}
\end{figure}

For the trend question, the bound is 0.34 to 0.43 in three representative cases
(Figure~\ref{fig:length}a, at 2026): a random walk with a cycle, a slowly reverting deviation with a
cycle, and a random walk alone. Even an ideal
test, told the exact alternative, would miss a true power law more often than it would detect it.
The question is answerable, but only just. For the reversion question the bound is 0.08 to 0.20,
for deviations reverting with half-lives from 0.55 to 2.2 years against non-reverting null models.
A test that ignored the data would reach 5\%, so whether deviations return to the trend is close
to unanswerable on this record. We therefore test the trend and report
only the bound for reversion. These bounds concern the six-feature summary and the evaluated
alternatives; they do not bound every possible use of the full series.

More data help the two questions very differently (Figure~\ref{fig:length};
Appendix~\ref{app:length} gives the numbers). We extended the record to 2030,
2036 and 2046 by simulation, under two assumptions about future volatility: that it stays at its
2026 level or keeps falling at its past rate. The number of low-frequency features grows with the
record, so that the fastest movement used keeps a period of about five years. The oracle bound for
the trend rises to 0.45--0.52 by 2030, 0.53--0.67 by 2036 and 0.64--0.86 by 2046. The bound for
reversion stays below 0.23 throughout. Our test's own power rises from about 0.26 today to
0.33--0.40 by 2030, 0.37--0.55 by 2036 and 0.46--0.75 by 2046. It captures a share of the oracle
gain that does not shrink as the record grows: 56--73\% today and 65--86\% by 2046. Even so, a
record more than twice as long would still miss a true power law about half the time in the hardest case.
Reversion would remain almost untouched.

\section{A test for a log-age trend under persistent deviations}\label{sec:method}

Our test compares two trends for expected log price: constant drift, and drift plus a log-age
term. It uses only the slow movements of the record, ignores the unknown size of the noise, and
takes the hardest deviation model in the declared family as its benchmark. Box~1 gives the recipe;
this section explains its parts. The derivations are in Appendix~\ref{app:model} and the calibration in
Appendix~\ref{app:calibration}.

\begin{figure}[!htb]
\centering
\fbox{\begin{minipage}{0.94\textwidth}\small
\textbf{Box 1. The test in brief.}
\begin{enumerate}\setlength\itemsep{1pt}
  \item Remove an intercept and a linear drift from the 820 weekly log prices.
  \item Summarise what remains by six weighted averages whose cosine weights complete between half
        a cycle and three cycles over the record. Call the result $z$.
  \item Keep only the direction of $z$, the unit vector $z/\lVert z\rVert$. Its distribution does
        not depend on the overall noise scale.
  \item Compute how likely this direction is under drift plus log age, with the log-age
        coefficient $b$ on a grid from 0 to 30, and under constant drift ($b = 0$). Maximise each
        over the 1,216 declared deviation models. The test statistic S1 is the difference of the
        two maximised log likelihoods.
  \item Reject constant drift at nominal level 5\% if $\mathrm{S1} > 2.632$. This critical value is
        the largest simulated 95\% quantile of S1 over the declared models, with an allowance for
        simulation error.
  \item For a lower confidence bound on $b$, subtract $b_0$ times the log-age signal from $z$ and
        repeat the test, for each $b_0$ on a fixed grid from $-30$ to 40 in steps of 0.1. The bound
        is the smallest $b_0$ not rejected.
\end{enumerate}
\end{minipage}}
\end{figure}

The first two steps follow \citet{muller2008}, who infer long-run properties of a series from a few
of its low-frequency averages. A trend over one decade of age lives at the lowest frequencies, but
so do persistent deviations, and the test has to weigh one against the other there. Faster
fluctuations add information about volatility but little about the trend. Six averages keep the
fastest movement used at a period of about five years, longer than the four-year cycle; with eight,
the fastest average would match the cycle's period (Appendix~\ref{app:calibration}). Twelve averages are a sensitivity
analysis.

The third step solves a problem specific to this setting. Under our deviation models, the variance
of weekly returns is not the variance that drives the slow movements, so the noise scale cannot be
estimated from returns and then plugged in. The direction of $z$ is unaffected by that scale: if
all deviations are multiplied by a constant, the direction does not change. Its density under a
Gaussian model with any mean and covariance can be computed exactly by a short recurrence
(Appendix~\ref{app:model}), so the likelihoods in step 4 involve no numerical approximation.

Steps 4 and 5 handle the breadth of the null hypothesis. Constant drift is not one model but 1,216:
every combination of persistence, cycle and volatility shape on the declared grid. A test must keep
its error rate below 5\% under each of them, so we take the likelihood under the most favourable
null model and calibrate against the hardest one. The critical value comes from simulating S1 under
every null member, refining the 60 hardest with 20{,}000 draws each, and taking the largest 95\%
quantile with a conservative allowance for simulation error. This is a numerical guarantee at
selected models, not a proof for every point of the continuous family. Section~\ref{sec:freeze} describes how we
checked it on fresh simulations before applying the test.

Step 6 turns the test into a statement about the size of the log-age term. If the true coefficient
is $b$, then removing $b$ times the log-age signal leaves data with no log-age term, and the test
rejects them with probability at most 5\% (nominally). The set of trial values not rejected is
therefore a confidence set, and its smallest member is a lower bound. This inversion is exact for
true values on the trial grid; between grid points it is an approximation.

Two secondary tests ask whether the power law itself needs repair, once constant drift is removed
from the alternatives. Both start from the residuals of a power law and look for an extra drift or a
curvature in log age, in either direction, with the same construction and calibration (critical
values 3.206 and 3.427). They are not confirmatory: a rejection would flag a departure, and a
non-rejection does not show that the power law is adequate.

Existing methods do not cover this setting. \citet{robinson2012} develops inference for trends of
the form $\beta t^\theta$ with stationary, short-memory errors. \citet{zhang2011} test a parametric
trend against a smooth alternative under short-range dependence. \citet{phillips2007} treats
regression on log age with stationary errors. Our deviations may be integrated, near-integrated or
fractional, which lies outside all three. The usual residual diagnostics behave poorly on this
design. In simulations of a power law with random-walk deviations, a residual unit-root test, as
commonly configured, rejected a unit root in 18--44\% of histories, and Newey--West intervals for
the exponent covered the truth in only 33--36\% (Appendix~\ref{app:conventional}). This is a property of those
procedures on this design, not of every possible calibrated alternative.

Beyond the declared family, the test is fairly robust (Appendix~\ref{app:stress}). With heavy-tailed or clustered
innovations, or a cycle whose period drifts, its rejection rate under constant drift stays between
2.5\% and 4.3\%. It rises to 6--10\% when volatility falls more steeply than the declared range
allows. The test also responds to other forms of
slowing growth: concave breaks in the drift are rejected in 7--14\% of histories. A rejection
therefore signals decelerating growth, of which the log-age term is one form.

\section{Pre-registration and validation}\label{sec:freeze}

A test is only as good as the discipline with which it is applied. In a reanalysis, the danger is
not fraud but flexibility: choices of data, window, model family and test, each reasonable, that
together steer the result. We removed that flexibility by fixing every choice before the test was
computed on Bitcoin data, and by recording the order of events.

We were not blind to this record. Two of our earlier papers analyse it. One, a preprint that has
not been peer reviewed, finds that an independent-errors likelihood comparison strongly favours the
power law over constant exponential growth \citep{baquero2026weak}. The other reports an in-sample
$R^2$ of 0.95 for the power law against 0.89 for constant exponential growth
\citep{baquero2026naive}. We also knew the published exponent and the residual diagnostics of
\citet{santostasi2026}. Both of our earlier comparisons point in the direction our test examines,
although neither accounts for persistent deviations. Some restrictions were also chosen after
simulation showed their effect on power: the cap on persistence at a random walk, and the range of
volatility shapes, whose upper end was then set by a bootstrap rule fixed before it ran. We
therefore present the study as a pre-registered reanalysis of a previously studied record. Its error
control refers to the fixed procedure under the declared models, not to an untouched sample.
Appendix~\ref{app:chronology} gives the full exposure statement.

Until the freeze, a written rule barred every computation on Bitcoin prices that could reveal the
answer: no trend test, no test of persistence or reversion, no curvature check. The only real-data
input to the simulations was the annual volatility of weekly returns, which is computed from
returns and carries no trend information. Every computation on real data, whatever its purpose,
was logged. The design was developed on simulated data and examined in five rounds of review before the
freeze. The reviewer was an AI model from a different provider than the one that assisted the
analysis (GPT-6-Astra, OpenAI; see the declaration at the end), so the review drew on a separately
trained system.

The protocol was frozen on 6 October 2026.\footnote{Twenty minutes after the freeze, at 17:29
UTC on 6 October 2026 and about thirteen hours before the test was run, the first author posted
the manifest's SHA-256 hash publicly
(\url{https://x.com/xmal/status/2107523626182308162}).} A manifest records the hash of every frozen file (code,
data, validation cells and protocol text), the exact constants, and one random seed for each
simulation, drawn from the operating system's entropy source. Validation then ran on these fresh
seeds, twice in separate processes, against six gates fixed in advance (Table~\ref{tab:gates}).
Each size gate asks whether the test keeps its false-rejection rate at most 6\%, with 95\%
simultaneous confidence, in 50 null models: the 20 hardest found in development, 20 structured
extremes and 10 drawn at random off the grid. The power gate asks whether the test reaches at least
half of the oracle bound's gain over 5\%. All six gates passed.

\begin{table}[t]
  \centering
  \caption{Validation gates, fixed before the freeze and run on fresh seeds. Each size gate uses
  50 null models with 20{,}000 simulated histories each; it passes if no model has more than 1{,}096
  false rejections (a one-sided Clopper--Pearson bound of 6\% at simultaneous confidence 95\%).}
  \label{tab:gates}
  \small
  \begin{tabular}{@{}lll@{}}
    \toprule
    Gate & What it checks & Result \\
    \midrule
    G1 & Size of S1 in 50 null models & largest rate 4.6\% (922 of 20{,}000) \\
    G5 & Size of the secondary tests (50 models each) & largest 4.9\% (drift), 4.7\% (curvature) \\
    G3 & Power at the published exponent (3 models) & 0.259, 0.263, 0.263 (targets 0.193, 0.204, 0.240) \\
    G4 & Coverage of the lower bound for $b$ & at least 0.949 (required 0.94) \\
    G4c & Implementation of the bound & no failure in 5{,}844 checks \\
    G6 & A second run reproduces the first & identical, bit for bit \\
    \bottomrule
  \end{tabular}
\end{table}

The test was then applied once, on 7 October 2026. Its outputs, including the forecast densities of
Section~\ref{sec:settle}, were archived with their hashes the same day. Later that day we registered the protocol,
the validation record, the results and the forecasts with the Open Science Framework
\citep{baquero2026osf}; it has been public since the working paper appeared. Because the registration followed the
run, it proves when the forecasts were fixed, before the prices they predict exist. For the
historical test, the evidence of order is the project's version history and the public post of the
manifest's hash before the run. The same reviewer then
reproduced every output from the frozen code and data. That audit found one reporting defect: two
output labels predated a wording change in the protocol. We corrected the labels in a dated note;
no number changed (Appendix~\ref{app:chronology}). Everything in this paper that uses Bitcoin prices outside the frozen
analysis, such as Figure~\ref{fig:apparent}a, is descriptive and labelled as such.

\section{Results}\label{sec:results}

The primary test rejects constant drift. Its statistic is $\mathrm{S1} = 3.410$, against a critical
value of 2.632 at the nominal 5\% level. The estimated maximum tail probability over the 50
validation models is 0.024: in the least favourable of them, about 2.4\% of simulated histories
without a log-age term give a statistic this large. Under the declared family of deviations, the
record therefore rejects constant drift in expected log price, in the direction of a positive
log-age component. The test's validated power at the published exponent was only about 26\% (Section~\ref{sec:freeze}). The rejection concerns the trend's growth rate, which Section~\ref{sec:method} showed responds to
other forms of deceleration as well. It is not, by itself, evidence for the power law.

The lower confidence bound for the log-age coefficient, on the frozen grid of trial values, is 1.5.
Every trial value from 1.5 up to the largest one, 10{,}000, is accepted, with no gap, so the
confidence set has no upper end on this grid. The rounded benchmark $b = 5.7$ is among the accepted
values. The bound concerns the log-age coefficient in the drift-and-log-age model, with the drift
left free; it is not an interval for the exponent of a pure power law. Its coverage is nominally
95\% for true values on the trial grid.

Neither secondary test detects a departure from the power law. Both statistics are exactly zero:
adding an extra drift or a curvature in log age does not improve the fit of the projected data
for any coefficient on the evaluated grid. Zero is a common outcome of these statistics, because
the grid of departures starts at a finite step: between 11\% and 31\% of the validation draws
without any departure are also exactly zero. The estimated tail probability is therefore one. This
means that neither prespecified departure was detected; it does not show that the power law is
adequate against every alternative.

\begin{figure}[t]
  \centering
  \includegraphics[width=0.92\textwidth]{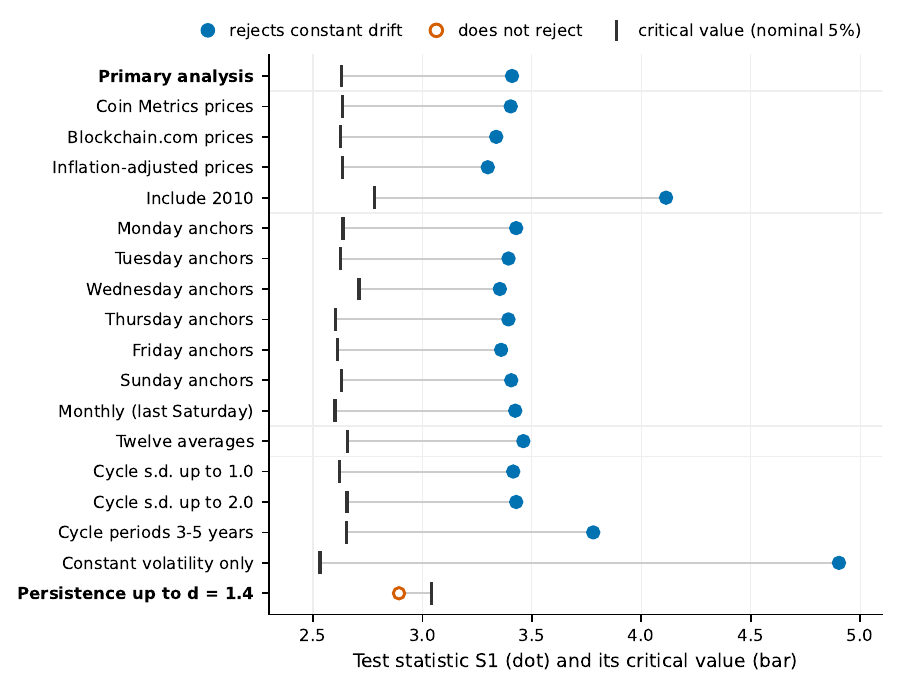}
  \caption{The primary test and its 17 sensitivity analyses. Each row shows the test statistic
  (dot) and that analysis's own critical value at the nominal 5\% level (bar), computed after the
  freeze and before the data were read. Filled dots reject constant drift. The analyses change the
  price source, the sampling, the number of low-frequency averages or the declared deviation
  family; they reuse substantially the same record, so they are not independent confirmations. The
  only non-rejection widens persistence beyond a random walk.}
  \label{fig:sensitivity}
\end{figure}

Figure~\ref{fig:sensitivity} shows the 17 sensitivity analyses fixed in the protocol
(Appendix~\ref{app:sensitivity} lists them). Changing the
price source, adjusting for inflation, starting in 2010, sampling on any other weekday or monthly,
using twelve averages, or changing the cycle's size or period all leave the decision unchanged; so
does restricting volatility to be constant. These analyses reuse substantially the same record, so
they show that the decision is not an artefact of a particular construction, not that it has been
confirmed sixteen times. One analysis does change the decision. When the deviation family is widened
to fractional persistence up to $d = 1.4$, beyond a random walk, the statistic falls to 2.893 against a
recalibrated critical value of 3.043. Such deviations are not larger but longer-lived: their changes
themselves persist, like momentum lasting years, so they trace smooth multi-year swings that can
mimic a bending trend (Table~\ref{tab:memory}). This is a different property from the decline in
volatility, which the primary family already includes. The primary rejection is therefore not robust to this
prespecified expansion of the persistence family. This does not show that Bitcoin's deviations are
that persistent. It shows that the conclusion rests on the assumption that they are not.

\begin{table}[t]
  \centering
  \caption{Descriptive Gaussian fits under the declared covariance family. Each trend model is
  fitted by maximum likelihood with the deviation model chosen from the declared grid. Differences
  in log likelihood are not calibrated tests, Bayes factors or model probabilities.}
  \label{tab:fits}
  \small
  \begin{tabular}{@{}lrl@{}}
    \toprule
    Trend model & $\Delta$ log likelihood & Fitted trend (descriptive) \\
    \midrule
    Drift and log age & $+0.22$ & log-age coefficient 6.57; drift $-0.08$ per year \\
    Gompertz & $+0.04$ & ceiling about 490{,}000 US dollars \\
    Power law & 0 & exponent 5.81 \\
    Logistic & $-1.93$ & ceiling about 79{,}000 US dollars; midpoint December 2021 \\
    Constant drift & $-7.19$ & growth 0.56 log units (75\%) per year \\
    \bottomrule
  \end{tabular}
\end{table}

The descriptive fits in Table~\ref{tab:fits} show how little the record separates the decelerating
curves. The power law, a Gompertz curve and the drift-and-log-age model have log likelihoods within
0.22 of one another; the logistic is about 2 units lower and constant drift about 7. The two models with
slightly higher likelihoods have one more parameter than the power law, and the drift-and-log-age
model contains it, so it must fit at least as well. These fits treat
the deviations as Gaussian with the declared covariance, so their differences are a description, not
a test. They do show that the record does not single out the power law: a curve with a ceiling near
490{,}000 dollars fits it as closely. The fitted exponent of the pure power law is 5.81. Finally,
measuring age from a later origin changes the fit only slightly: the profile over shifts of the
origin, at the covariance of the genesis fit, peaks 32 weeks after the genesis block with a gain of
only 0.19 in log likelihood. The genesis block is a reasonable origin, but the data do not pin one
down.

\section{Prospective forecasts and longer records}\label{sec:settle}

The record cannot single out the shape of the trend, but future prices can. Before any of them
exist, we fixed and registered a forecast from each of the five trend models
\citep{baquero2026osf}. Each forecast is a predictive density for log price on the last Saturday of
2027, 2028, 2029 and 2030. It mixes the conditional distribution at the fitted parameters with 49
others from a parametric bootstrap, so that some of the uncertainty in the fit is carried forward.
When the four prices are known, each model will be scored by the log density it assigned to them
jointly, with the four separate scores as a secondary measure. A higher score means better
prediction of these four prices, not proof of a generating law.

\begin{figure}[t]
  \centering
  \includegraphics[width=\textwidth]{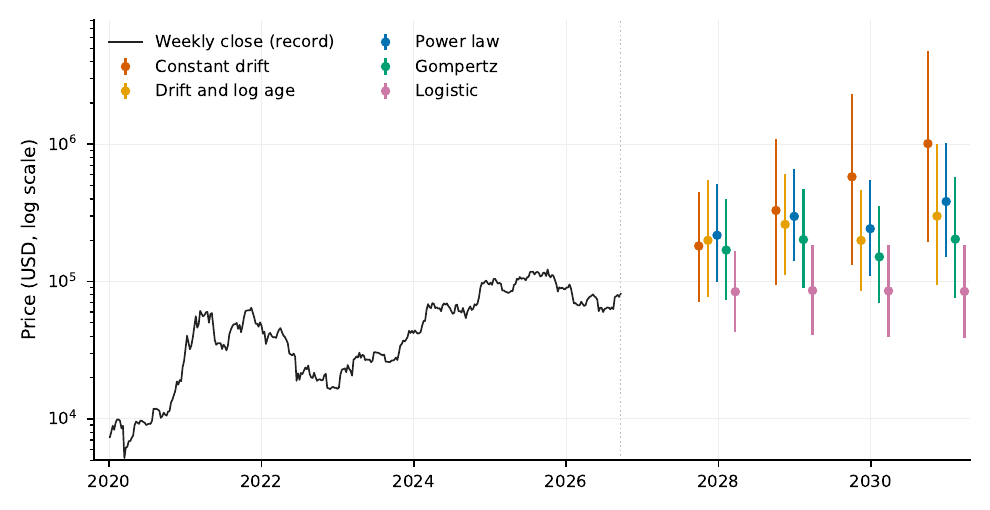}
  \caption{The registered forecasts. Weekly closing price since 2020 (black line; the record ends
  at the dotted line), and for each trend model the median (dot) and 90\% interval (bar) of its
  predictive density at the four target dates: the last Saturday of 2027, 2028, 2029 and 2030.
  Within each year the models are offset horizontally for legibility. The densities were fixed
  and registered on 7 October 2026 \citep{baquero2026osf}.}
  \label{fig:forecast}
\end{figure}

Figure~\ref{fig:forecast} and Table~\ref{tab:forecast} show the forecasts. By the end of 2030 their medians range from about
85{,}000 dollars under the logistic curve to about 1{,}000{,}000 under constant drift, with the
Gompertz curve (about 200{,}000), the drift-and-log-age model (about 300{,}000) and the power law
(about 380{,}000) in between. The intervals overlap widely: the power law's 90\% interval runs
from about 150{,}000 to 1{,}000{,}000 dollars. The dip in 2029 under the power law, the drift-and-log-age model
and the Gompertz curve comes from the stochastic cycle in their fitted deviation models, with a
period of 3.5 years, projected forward; the other two fits have no cycle. One year's price will therefore separate the models
only weakly. Four years, scored jointly, should separate constant drift and the logistic curve
from the rest more clearly than the past record can. The three trends that fit the record best will
take longer (Appendix~\ref{app:discrimination}): the power law and the Gompertz curve should
separate in the mid-2030s, while a power law with a small added drift may never be distinguished
from a pure one.

\begin{table}[t]
  \centering
  \caption{The registered forecasts: median and 90\% interval of each model's predictive density
  for the closing price on the last Saturday of each year, in thousands of US dollars. The record
  ends at about 81 thousand dollars (19 September 2026). These are model-dependent predictive
  densities, probably too narrow (see text), not calibrated intervals.}
  \label{tab:forecast}
  \small
  \begin{tabular}{@{}lllll@{}}
    \toprule
    Model & 25 Dec 2027 & 30 Dec 2028 & 29 Dec 2029 & 28 Dec 2030 \\
    \midrule
    Constant drift & 181 (70--449) & 329 (94--1{,}087) & 579 (133--2{,}340) & 1{,}011 (194--4{,}791) \\
    Drift and log age & 200 (77--552) & 261 (111--603) & 199 (86--464) & 300 (95--1{,}002) \\
    Power law & 217 (99--511) & 298 (140--662) & 243 (111--547) & 383 (150--1{,}015) \\
    Gompertz & 170 (73--397) & 202 (90--471) & 151 (69--353) & 204 (76--579) \\
    Logistic & 84 (43--166) & 86 (41--186) & 85 (40--184) & 85 (39--183) \\
    \bottomrule
  \end{tabular}
\end{table}

These forecasts depend on their models in a way the reader should see. The record ends at about
81{,}000 dollars, below four of the five fitted trends (about 104{,}000 to 130{,}000 dollars; the
logistic trend is at 79{,}000). The deviation models selected in
the fits revert towards the trend, with a half-life of about half a year for three of the five, so
most forecasts expect the price to rise towards the trend. Reversion is precisely the property
the record cannot identify (Section~\ref{sec:identifiability}). The scores will therefore reflect the assumed reversion as
well as the shape of the trend. The forecasts are also likely too narrow. \citet{muller2016} show
that when the persistence of a series is uncertain, long-run predictive sets that account for that
uncertainty are wider than those built with persistence fixed, and that ignoring it understates
forecast uncertainty. Our bootstrap varies the deviation model only among neighbouring members of
the declared grid. We therefore present these as model-dependent predictive densities, not as
calibrated intervals.

A longer record will also sharpen the test itself (Figure~\ref{fig:length}). With the same rule,
our test's power at the published exponent would rise from about 0.26 today to roughly 0.4--0.55
by 2036 and 0.46--0.75 by 2046. Repeating the analysis on a longer record would need its own
protocol and calibration, but no new ideas. The reversion question will remain hard: even in 2046,
no test using the same information could detect the evaluated reverting deviations more than about
a quarter of the time.

A different route is to bring in information other than time. The mechanism proposed by
\citet{santostasi2026} links price to the growth of addresses and to the size of the network, which
suggests testing price against on-chain measures directly. Such a study faces its own pitfalls.
Regressions between persistent series can show strong relations where none exist
\citep{phillips1986}, and measures that depend on price, such as values denominated in dollars, can
carry price information into the regressor. It would need its own protocol, firewall and
calibration. Our test uses time since the genesis block as its only regressor, and its result says
nothing about whether on-chain measures explain price better.

\section{Discussion}\label{sec:discussion}

The record supports one claim and leaves the others open. Under a broad, declared family of
persistent deviations, constant drift in expected log price is rejected in the direction of a
positive log-age component: Bitcoin's trend growth has slowed in the way a power law in time would
imply. On the frozen grid of trial values, the lower confidence bound for the log-age coefficient is 1.5,
and the rounded published value of 5.7 is among the values the data accept. In this sense the trend part of the
claim is consistent with the record, within the declared family.

Three things remain open, and the paper's title names the main one. The data do not single out
the power law: a Gompertz curve, which grows towards a ceiling of about half a million dollars, and
a power law with added drift describe the record as well. Whether deviations return to the trend
cannot be decided from fifteen years of data, so we neither support nor refute that part of the
claim. And the rejection itself rests on capping persistence at a random walk: allowing more
persistent deviations, as one of our prespecified sensitivity analyses does, removes it. None of
this concerns the mechanism proposed for the power law, which time alone cannot test.

The result is conditional in further ways. Error control holds for the declared models, at the
nominal 5\% level, and was checked numerically at selected models with a 6\% tolerance; it is not
a proof over the whole continuous family. The volatility profile is a one-parameter family; when
volatility falls more steeply than it allows, the false-rejection rate rises to 6--10\%. The
descriptive fits and the forecasts assume Gaussian deviations with the declared covariance. The
price series splices two exchanges at the start of 2013, although the conclusion survives other
sources. And the record had been studied before, including by us, so the study is a reanalysis
with a fixed procedure, not a fresh confirmation.

Our findings sit alongside those of \citet{santostasi2026} rather than against them. We agree that
growth has decelerated and that an exponent near their estimate is consistent with the data. We
differ on what the record can establish about mean reversion. Their published conclusion of
trend-stationarity \citep{santostasi2026} rests on residual unit-root and stationarity tests; in our simulations, such tests, as commonly configured, do not
keep their nominal properties when the deviations are as persistent as these (Appendix~\ref{app:conventional}). That is
a property of the procedures on this design, not a judgement on the claim, which the record simply
cannot decide.

A concurrent analysis by the same authors \citep{santostasi2026clauset}, deposited on
2 October 2026, takes the power law as the null hypothesis and asks whether the record is consistent
with it, using skew-normal residuals and synthetic histories with stationary autoregressive deviations
matched to the data. The two studies ask complementary questions, and they
agree where they overlap: neither detects curvature against the power law, both find an exponent near
the published value, and both conclude that the record cannot separate the power law from rival growth
laws. They found a dependence-corrected comparison with exponential growth uninformative; our test,
built for that comparison under persistent deviations, rejects constant drift within the declared
family. Neither analysis supports a conclusion about reversion to the trend.

The approach carries beyond Bitcoin. Claims of power-law growth in time are made for other young,
fast-growing systems, often on short records and with a mechanism attached. Confirmed COVID-19 cases
in Chinese provinces grew roughly as $t^{2.1}$ over about three weeks in early 2020, a pattern
attributed to containment \citep{maier2020}; early epidemic growth is usually slower than exponential,
often polynomial in time, in diseases from HIV/AIDS to Ebola \citep{chowell2016}. Each such claim faces
the question we faced here: how much can a short record say about the shape of its trend? For such claims the order of steps matters. First ask what the record could reveal
under the best possible test, given a declared family of deviations. Then test only what is
answerable, with a procedure calibrated over that family. Fix the procedure before looking, and
publish its hash. Finally, register forecasts, because future observations can settle what the
past cannot. Each step is cheap; together they turn an appealing line on a log--log plot into a
statement whose strength and limits are both visible.

\section*{Data availability}
The code, the frozen protocol and its manifest, the validation and analysis records, the derived
statistics behind every confirmatory result, and the review record are deposited at Zenodo
(replication package, \href{https://doi.org/10.5281/zenodo.23218897}{doi:10.5281/zenodo.23218897}). Price data are not redistributed: a script in the deposit downloads them from
their public sources and rebuilds the analysed series byte for byte. The protocol, results and
forecasts are registered at the Open Science Framework \citep{baquero2026osf}.

\section*{Declaration on the use of AI}
The analysis code, simulations and draft text were prepared with the assistance of Claude (Opus 5.5,
Anthropic), under the direction of the authors. The design, code and results were reviewed by a
second AI model, GPT-6-Astra (OpenAI). The authors take full responsibility for the content.

\bibliographystyle{abbrvnat}
\bibliography{refs}

\clearpage
\appendix
\renewcommand{\thesection}{\Alph{section}}
\begin{center}\Large\bfseries Supplementary appendices\end{center}

\section{Data}\label{app:data}

\paragraph{Primary series.} Daily closing prices in US dollars, one per UTC day, built by
\nolinkurl{data/make_exchange_series.py} from hashed raw records. Until 31 December 2012 the close is
the last Mt.\ Gox trade of each UTC day, from the complete Mt.\ Gox trade archive collected by
bitcoincharts.com and preserved by the Internet Archive. From 1 January 2013 it is Bitstamp's
official daily close, which we checked against Bitstamp's archived trades from 1 January 2013 to
9 July 2014. The levels of the two exchanges differ by 0.6\% at the splice, and we keep that gap. The
series file has SHA-256 \texttt{\seqsplit{5552aa23e37dd5d972a5956c44aa6682a7aa60ab2fe3bbf06e99a9e5023dacf7}}.
We use closes rather than daily averages, because averaging induces spurious autocorrelation in
returns, which would contaminate the persistence question.

\paragraph{Observation design.} Every Saturday from 1 January 2011 to 19 September 2026 is an
anchor; each value is that day's close, timed at the end of the UTC day. One anchor, 25 June 2011,
has no close and is dropped, leaving 820 observations. Age is measured in weeks from the genesis
block (3 January 2009, 18:15:05 UTC). The model runs on a weekly latent grid from the genesis week,
and the same observation operator (keep the observed anchors) applies to means, covariances and
simulations.

\paragraph{Sensitivity series.} \nolinkurl{data/sensitivity_daily_2026-10-05.csv} (SHA-256
\texttt{\seqsplit{4113e284f579a284648f5a792cfd5fcb2171c88e85ef02fef1d5b9b6e797d50b}}) holds the Coin Metrics and
Blockchain.com reference prices, the primary series from July 2010, and the primary series deflated
by the US consumer price index (FRED series CPIAUCNS). Each source's missing anchors come from its
own availability.

\paragraph{Volatility profile.} The only real-data input to the simulations before the freeze is the
annual standard deviation of weekly log returns, computed from returns alone:

\begin{center}\small
\begin{tabular}{@{}*{8}{c}@{}}
\toprule
2011 & 2012 & 2013 & 2014 & 2015 & 2016 & 2017 & 2018 \\
0.294 & 0.093 & 0.212 & 0.089 & 0.089 & 0.063 & 0.140 & 0.134 \\
\midrule
2019 & 2020 & 2021 & 2022 & 2023 & 2024 & 2025 & 2026 \\
0.085 & 0.109 & 0.111 & 0.092 & 0.071 & 0.065 & 0.051 & 0.065 \\
\bottomrule
\end{tabular}\end{center}

\section{Model and statistics}\label{app:model}

\paragraph{Deviation models.} Deviations $u$ live on the weekly latent grid $k = 0, 1, \dots$ from the
genesis week and are the sum of one persistence component and one cycle, independent of each other.
Innovations $e_k$ have standard deviation $\sigma_k = \bar\sigma\, h_k^\lambda / \mathrm{rms}(h^\lambda)$,
where $h_k$ is the annual volatility of Appendix~\ref{app:data} for the calendar year of week $k$ (2011's value
before 2011), the root mean square is taken over a fixed calibration support, and $\bar\sigma =
0.128$ is a reference scale that the statistic does not depend on. Persistence states start at zero
in the genesis week:
\begin{itemize}\setlength\itemsep{1pt}
  \item autoregressive: $u_k = \rho u_{k-1} + e_k$, $\rho = 1 - c/821$, $c \in \{0, 1, 2.5, 5, 10, 20, 40\}$
        ($c = 0$ is the random walk);
  \item fractional (type II from genesis): $u_k = \sum_{j \le k} \psi_{k-j}(d) e_j$ with
        $\psi(d)$ the coefficients of $(1-L)^{-d}$, $d \in \{0.25, 0.4, 0.5, 0.6, 0.75, 0.9\}$;
  \item local level: $u_k = \epsilon_k + g\, w_k$, $w_k = \sum_{j \le k} \eta_j$, with $\epsilon$ and
        $\eta$ independent, of standard deviation $\sigma_k$, $g \in \{0.1, 0.3, 0.5\}$.
\end{itemize}
The cycle is Harvey's stochastic trigonometric cycle, stationary from genesis, with covariance
$\mathrm{Cov}(\psi_k, \psi_l) = (r\bar\sigma)^2 \rho_c^{|k-l|} \cos\{\omega(k-l)\}$, where
$\omega = 2\pi/P$ for a period of $P$ weeks and $\rho_c = 0.5^{1/(\kappa P)}$, so that the correlation
envelope halves after $\kappa$ cycles. The grid is $r \in \{0, 5.86, 11.72\}$ (a standard deviation of
up to 1.5 log units), $P \in \{3.5, 4, 4.5\}$ years and $\kappa \in \{1, 3, 5\}$ cycles; together with
$\lambda \in \{0, 0.5, 1, 1.25\}$ this gives $16 \times 19 \times 4 = 1{,}216$ members.

\paragraph{Projections.} Let $M_X$ residualise on the columns of $X$, and let $\Psi$ hold the cosine
weights $\Psi_{tj} = \sqrt{2} \cos(\pi j \tau_t)$, $j = 1, \dots, q$, where $\tau_t$ is the position of
anchor $t$ in the record. For S1, $z = \Psi^\top M_{[1, s]}\, y$, with $q = 6$. Under deviation model
$\theta$ and trend coefficient $b$, $z \sim N(b\,m, \Sigma_\theta)$ with $m = \Psi^\top M_{[1,s]} \log s$ and
$\Sigma_\theta$ computed exactly from the latent covariance.

\paragraph{Direction density.} For a unit vector $u = z/\lVert z\rVert$, write
$a = u^\top \Sigma^{-1} u$, $g = u^\top \Sigma^{-1} m$ and $c = m^\top \Sigma^{-1} m$. Up to a constant common
to all models,
\[
  \log f(u; b m, \Sigma) = -\tfrac12 \log|\Sigma| - \tfrac12 b^2 c - \tfrac{q}{2} \log a
  + \log J_{q-1}\!\left(b g / \sqrt{a}\right),
  \qquad J_n(x) = \int_0^\infty t^n e^{-t^2/2 + x t}\, dt .
\]
$J_n$ follows the exact recurrence $J_0(x) = \sqrt{\pi/2}\, e^{x^2/2} \operatorname{erfc}(-x/\sqrt2)$,
$J_1 = 1 + x J_0$, $J_n = (n-1) J_{n-2} + x J_{n-1}$, evaluated in the scaled form $J_n e^{-x^2/2}$,
which is stable for $x \ge 0$. If $g \le 0$, both $b$-dependent terms are non-increasing in $b \ge 0$,
so the maximum over $b \ge 0$ is at $b = 0$.

\paragraph{The statistics.} With $\Theta$ the 1{,}216-member grid and $B = \{0, 0.75, \dots, 30\}$
at the reference scale,
\[
  \mathrm{S1}(u) = \max_{\theta \in \Theta,\, b \in B} \log f(u; b m, \Sigma_\theta)
                 - \max_{\theta \in \Theta} \log f(u; 0, \Sigma_\theta) \;\ge\; 0 .
\]
A zero projected vector, or any non-finite value, stops the computation. The secondary tests use
$z = \Psi^\top M_{[1, \log s]}\, y$ and a departure regressor $v$: extra drift ($v = s$) or curvature
($v = (\log s - \overline{\log s})^2$), each scaled to unit range after removing its best power-law
fit, with coefficients in $\pm B$. The two-sided maximum uses the symmetry $f(u; -bm, \Sigma) =
f(-u; bm, \Sigma)$.

\section{Calibration and validation}\label{app:calibration}

\paragraph{Critical values.} For each statistic, every member of $\Theta$ was simulated 2{,}000 times
under the null; the 60 members with the highest 95\% quantiles were refined with 20{,}000 draws each;
the critical value is the largest refined 97.5\% order-statistic upper confidence limit for the 95\%
quantile. This gave 2.631683520034235 for S1 (seed 20261024; largest refined quantile 2.548),
3.20558780703057 for the drift test and 3.4266230049087683 for the curvature test (seeds 20261027 and
20261028, with members flagged by an earlier pilot added to the refinement). In development, the
largest rejection rate of S1 at 40 random off-grid members (4{,}000 draws each) was 4.5\%.

\paragraph{Six averages.} With $q = 6$, the fastest cosine has a period of about 5.2 years. With eight,
the fastest average would match the four-year cycle and, in development simulations, bias a
persistence test towards reversion. Twelve averages are a sensitivity analysis.

\paragraph{The volatility range.} A first family allowed $\lambda$ up to 1; a stress test showed that
steeper declines inflated the rejection rate, and widening to $\lambda \le 2$ restored size but cost
power at the published exponent below the design target. Following that failure, we adopted a narrower
restriction whose upper end was set by a rule fixed before it ran: the 99th percentile of a four-week
block bootstrap of the steepness of the annual volatility profile, which gave 1.24, rounded up to
1.25. This rule measures uncertainty in the profile's overall steepness; it is not a confidence region
for the full volatility profile, and error control is conditional on the declared family. The failed
wider-family run is part of the record.

\paragraph{Validation.} Gates and cells were fixed in the frozen manifest; validation ran on seeds
drawn at the freeze (Table~\ref{tab:gates}). The 50 cells for S1 are its 20 hardest members in
development, 20 structured extremes (persistence $\{c = 0, c = 40, d = 0.25, g = 0.5\}$ crossed with no
cycle or the largest cycle, and $\lambda \in \{0, 1, 1.25\}$) and 10 off-grid members drawn at random;
each secondary test has its own 20 hardest members and shares the other 30. Each statistic forms its
own family of 50 cells with 95\% simultaneous confidence; no joint statement over the 150 cells is
made. The power gate's targets are $0.05 + 0.5(B - 0.05)$, half of the oracle bound's gain over 5\%.

\section{The lower confidence bound}\label{app:bound}

The trial set $T$ has 718 points: $-30$ to 40 in steps of 0.1 (rounded to ten decimals, so that 0 and
5.7 are members), extended by doubling to $\pm 10{,}000$. The confidence set is $C_T(z) = \{t \in T :
\mathrm{S1}(z - t m) \le c\}$ and the bound is $L_T(z) = \min C_T(z)$, or $+\infty$ if $C_T$ is
empty. For any true $b \in T$, $z - bm$ has a null distribution, so $\mathrm{S1}(z - bm) \le c$ puts $b$
in $C_T$ and hence $L_T \le b$: $P(L_T \le b) \ge 1 - P\{\mathrm{S1}(z - bm) > c\}$, with no assumption
about the shape of $C_T$. For true values between trial points no coverage is claimed. A set accepted
only between two neighbouring trial points is not seen: for example, if the acceptance set were
$[-9.976, -9.974] \cup [3, \infty)$, the bound would be 3. Flags record an empty set, acceptance at the
lowest trial point, a rejected trial point above $L_T$ (disconnected set), and a zero recentred
vector. The validation check (gate G4c) used 2{,}000 histories at $b = 5.7$ in each of three models;
in all 5{,}844 histories where S1 accepted 5.7, the bound was at most 5.7, and no bound was undefined.

\section{Behaviour beyond the declared family}\label{app:stress}

Rejection rates of S1 under constant drift, at its frozen critical value, with 4{,}000 simulated
histories per scenario:

\begin{center}\small
\begin{tabular}{@{}lr@{}}
\toprule
Scenario & Rejection rate \\
\midrule
Inside the family: random walk, annual volatility profile & 3.9\% \\
Inside the family: AR ($c = 5$) with cycle & 2.4\% \\
Heavy tails ($t_4$), GARCH(1,1), stochastic volatility, cycle with drifting period & 2.5--4.3\% \\
Volatility shape $\lambda = 1.5$ / $\lambda = 2$ & 4.8\% / 6.1\% \\
Extra pre/post-2014 volatility tilt of 25\%, 50\%, 100\% & 5.4\%, 7.7\%, 9.7\% \\
Twofold volatility break, random walk & 9.9\% \\
\midrule
Concave breaks in the drift (decelerating) & 7.0--13.9\% \\
Convex break in the drift & 1.5\% \\
Level jump & 3.8\% \\
\bottomrule
\end{tabular}\end{center}

The first block measures size; the second shows that S1 responds to deceleration in other forms than
a log-age term.

\section{Sensitivity analyses}\label{app:sensitivity}

Each analysis changes one element of the primary design; its critical value was computed after the
freeze by the same rule, with its own seed, before any price was read, and its observed dates were
checked against those of the calibration.

\begin{center}\footnotesize
\begin{tabular}{@{}llrrrr@{}}
\toprule
Analysis & Change & Models & Obs. & Critical & S1 \\
\midrule
Primary & --- & 1216 & 820 & 2.632 & 3.410 \\
Coin Metrics & price source & 1216 & 821 & 2.634 & 3.404 \\
Blockchain.com & price source & 1216 & 821 & 2.626 & 3.337 \\
Inflation-adjusted & prices deflated by CPI & 1216 & 820 & 2.636 & 3.299 \\
Include 2010 & start 17 July 2010 & 1216 & 844 & 2.780 & 4.113 \\
Monday \ldots\ Sunday & anchor weekday & 1216 & 819--821 & 2.602--2.711 & 3.354--3.428 \\
Monthly & last Saturday of each month & 1216 & 187 & 2.601 & 3.424 \\
Twelve averages & $q = 12$ & 1216 & 820 & 2.658 & 3.461 \\
Cycle up to 1.0 & cycle s.d.\ at most 1.0 log units & 1216 & 820 & 2.621 & 3.415 \\
Cycle up to 2.0 & cycle s.d.\ at most 2.0 log units & 1216 & 820 & 2.656 & 3.429 \\
Wide cycle & periods 3--5 y, coherence $\ge 0.5$ & 1600 & 820 & 2.653 & 3.781 \\
Constant volatility & $\lambda = 0$ only & 304 & 820 & 2.532 & 4.903 \\
Persistence to $d = 1.4$ & adds $d \in \{1.1, 1.25, 1.4\}$ & 1444 & 820 & 3.043 & 2.893 \\
\bottomrule
\end{tabular}\end{center}

The six weekday analyses end on their own last weekday before 25 September 2026 (18 to 24 September),
so they change the endpoint as well as the phase. The monthly design schedules the calendar month's
last Saturday, excludes the incomplete September 2026, and drops a missing date rather than
replacing it (June 2011 is dropped).

\section{Conventional diagnostics on this design}\label{app:conventional}

To show why we calibrate by simulation, we applied common residual diagnostics to simulated daily
histories with a true power law (exponent 5.7) plus deviations of six kinds, 500 histories each. The
procedures follow common practice: least squares on $[1, \log t]$; a residual augmented Dickey--Fuller
test (constant, lag length by AIC, MacKinnon critical values) and a KPSS test (level, automatic lags),
each at 5\%; Newey--West standard errors with a four-year window; and a check that every point from
2021 to 2026 lies within $\pm 2\sigma$ of a fit to 2020.

\begin{center}\small
\begin{tabular}{@{}lccccc@{}}
\toprule
Deviations & ADF rejects & KPSS accepts & Both & NW coverage & In band \\
\midrule
Random walk, volatility profile & 0.44 & 0.06 & 0.05 & 0.35 & 0.32 \\
Random walk, constant volatility & 0.18 & 0.04 & 0.03 & 0.34 & 0.09 \\
Random walk + cycle, profile & 0.20 & 0.15 & 0.04 & 0.36 & 0.24 \\
Random walk + cycle, constant & 0.06 & 0.12 & 0.01 & 0.33 & 0.10 \\
AR half-life 2.2 years + cycle & 0.20 & 0.20 & 0.06 & 0.44 & 0.38 \\
AR half-life 0.55 years + cycle & 0.31 & 0.60 & 0.19 & 0.62 & 0.44 \\
\bottomrule
\end{tabular}\end{center}

The first four rows have no reversion, so any rejection of a unit root is an error; the residual ADF
test alone makes it in 18--44\% of histories, mainly because declining volatility makes early
deviations large and late ones small. Requiring both an ADF rejection and a KPSS non-rejection
rarely errs (1--5\%) but also rarely detects true reversion (6\% and 19\%). Newey--West intervals for
the exponent cover the truth in 33--62\% of histories against a nominal 95\%. These are properties of
the procedures as configured on this design, not of every calibrated alternative.

\section{Descriptive fits and forecasts}\label{app:fits}

\paragraph{Fits.} Each trend model of Table~\ref{tab:trends} is fitted by Gaussian maximum likelihood on
the 820 observations, with covariance $\kappa^2 \Omega_\theta$, $\theta$ on the declared grid and
$\kappa^2$ profiled out. Linear models are profiled exactly over $\theta$. The logistic and Gompertz
curves use a fixed two-stage search: bounded quasi-Newton fits of their nonlinear parameters from
fixed starting points at 24 evenly spaced grid members; with the best of these fixed, a profile over
all members; then a refit at the selected member, kept only if it does not lower the likelihood.
Only optimiser runs that report success count. We compare these models descriptively because the
logistic and Gompertz families are not closed under rescaling, so the direction-based construction of
Section~\ref{sec:method} cannot remove the noise scale for them.

\paragraph{Origin profile.} The power law is refitted with age measured from the genesis block plus a
shift $\tau$, from $-104$ to $+104$ weeks in steps of 8, holding the deviation model at that of the
genesis fit and refitting the intercept, slope and scale. It is a conditional profile, not a test or a
confidence set for the origin.

\paragraph{Forecasts.} For each model, the predictive density of log price at the four target dates is
an equal-weight mixture of 50 Gaussians, each the distribution of the targets given the whole observed
path: one at the fitted parameters and 49 from a parametric bootstrap. Each bootstrap history is
simulated from the fitted model and refitted with the deviation model restricted to the fitted member
and its neighbours on the grid (one step in exactly one coordinate); its conditional distribution is
then evaluated on the actual path. Future volatility stays at its 2026 value. All five models had 49
successful refits. The densities, with their hash, are in the registration \citep{baquero2026osf}.

\section{Chronology and exposure}\label{app:chronology}

\paragraph{Order of events.}
\begin{center}\small
\begin{tabularx}{\textwidth}{@{}lX@{}}
\toprule
Date (2026) & Event \\
\midrule
25 Sept -- 5 Oct & Data sources built; identifiability maps; design developed on simulation; review rounds 1--3 \\
2 Oct & Concurrent analysis deposited \citep{santostasi2026clauset}; known, deliberately not read \\
6 Oct, 17:09 UTC & Protocol frozen; manifest SHA-256 \texttt{3d1884dd\ldots5882a} (after reviews 4--5) \\
6 Oct, 17:29 UTC & Manifest hash posted publicly \\
6--7 Oct & Validation (two runs) and sensitivity calibrations; receipt: all gates pass \\
7 Oct, 06:43 UTC & Single application to Bitcoin prices \\
7 Oct, 09:23 UTC & Forecasts and results registered (osf.io/2rdty); made public the same day with the working paper \\
7 Oct & Reproduction audit (review 6); reporting correction 1 \\
7 Oct, 14:59 UTC & Concurrent analysis first read \\
\bottomrule
\end{tabularx}\end{center}

\paragraph{What we knew before the freeze.} No test of trend shape, persistence or reversion was run on
Bitcoin prices before the freeze; every real-data computation was logged. We knew: the least-squares
exponent of our earlier study \citep[5.64;][]{baquero2026weak} and that study's comparisons, including
an independent-errors likelihood strongly favouring the power law over constant exponential growth and
a first-order autocorrelation of residuals close to one; the in-sample comparison of
\citet{baquero2026naive} ($R^2$ 0.95 against 0.89); the published exponent, residual diagnostics and
trend-stationarity conclusion of \citet{santostasi2026}; published unit-root and cointegration results
in a related post \citep{perrenod2026}; and the general halving-cycle pattern of Bitcoin prices. A concurrent analysis by
\citet{santostasi2026clauset} was deposited on 2 October 2026, four days before the freeze. Its authors
told the first author of it that day, and he replied that he would not read it until this work was
done; it was first read on 7 October 2026 at 14:59 UTC, after the single run and the registration. The
signal size $b = 5.7$ used throughout the simulations comes from the published exponent. The cap on
persistence and the volatility range were chosen after simulations of attainable power, as described
in Section~\ref{sec:freeze} and Appendix~\ref{app:calibration}.

\paragraph{Reviews and corrections.} The design and code were reviewed in five rounds before the freeze
and audited after the run by an AI model from a different provider (GPT-6-Astra, OpenAI); each review
and reply is archived with the artifact. The audit reproduced every output. It found that two output
labels predated a wording change in the protocol: the tail-probability label and the lower bound's
label. Reporting correction 1 records the original and corrected labels; no number, decision or
forecast changed.

\section{Power against record length}\label{app:length}

Oracle upper bound and achieved power of S1 at $b = 5.7$, for records ending at four dates and two
assumptions about volatility after 2026 (flat at the 2026 level, or declining at its past rate). The
number of averages grows with the record ($q = 6, 8, 10, 14$) so that the fastest keeps a period of
about five years, and each record has its own critical value by the frozen rule. This study is
exploratory and not part of the protocol.

\begin{center}\small
\begin{tabular}{@{}lccc@{}}
\toprule
Record ends & RW + cycle, $\lambda = 1$ & AR ($c = 5$) + cycle, $\lambda = 1$ & RW, $\lambda = 0$ \\
\midrule
2026 & 0.26 (0.34) & 0.26 (0.36) & 0.26 (0.43) \\
2030 flat / decline & 0.35 / 0.36 (0.45 / 0.46) & 0.38 / 0.40 (0.49 / 0.51) & 0.33 / 0.34 (0.51 / 0.52) \\
2036 flat / decline & 0.41 / 0.45 (0.53 / 0.57) & 0.50 / 0.55 (0.64 / 0.67) & 0.37 / 0.40 (0.58 / 0.60) \\
2046 flat / decline & 0.53 / 0.59 (0.64 / 0.75) & 0.68 / 0.75 (0.81 / 0.86) & 0.46 / 0.50 (0.66 / 0.75) \\
\bottomrule
\end{tabular}\end{center}

Entries are achieved power with the oracle bound in parentheses. Oracle bounds for reversion (AR
deviations with half-lives of 0.55 to 2.2 years, with a cycle) rise from 0.09--0.12 in 2026 to
0.16--0.22 in 2046.

\section{Separating the remaining trends with future data}\label{app:discrimination}

The power law, the drift-and-log-age model and the Gompertz curve describe the record almost equally
well (Table~\ref{tab:fits}). To see when future prices could separate them, we treated each in turn
as true and simulated 1{,}000 continuations of the record, weekly to the end of 2046, from its
predictive distribution given the observed prices, at its fitted parameters. All three fits select
the same deviation model (a near-random walk with a 3.5-year cycle); we held it fixed and known,
with volatility kept at its 2026 level. At record ends in 2030, 2035, 2040 and 2046 we refitted the
three trends and compared them by log-likelihood penalised by the number of trend parameters
(Akaike's criterion). We call a comparison decisive when the true model leads a rival by at least 3,
odds of about 20 to 1. This study is exploratory and not part of the protocol; it uses the
fitted parameters, so it is post hoc.

\begin{center}\small
\begin{tabular}{@{}llcccc@{}}
\toprule
True model & Rival & 2030 & 2035 & 2040 & 2046 \\
\midrule
Power law & Gompertz & 0.14 & 0.75 & 0.93 & 0.98 \\
Gompertz & Power law & 0.16 & 0.98 & 1.00 & 1.00 \\
Drift and log age & Power law & 0.00 & 0.02 & 0.19 & 0.71 \\
Drift and log age & Gompertz & 0.00 & 0.08 & 0.37 & 0.72 \\
Power law & Drift and log age & 0.00 & 0.00 & 0.00 & 0.00 \\
Gompertz & Drift and log age & 0.00 & 0.00 & 0.00 & 0.00 \\
\bottomrule
\end{tabular}\end{center}

Entries are the proportion of simulated records in which the true model is decisively ahead of the
rival; the rival was never decisively ahead. Three patterns stand out. The power law and the
Gompertz curve separate in the mid-2030s, sooner if the price levels off. The drift-and-log-age
model contains the power law, so when the power law is true the extra drift can only be shown to be
small, never absent; when the fitted drift (about $-8\%$ a year) is real, it becomes detectable in
the 2040s. And a power law with a small negative drift bends much as a curve approaching a ceiling
does, so it cannot be ruled out when the Gompertz curve is true, even by 2046. Holding the deviation
model known makes these proportions optimistic: estimating it, as a real analysis must, would slow
the separation.

\end{document}